\documentclass[11pt,a4paper]{article}
\usepackage{jheppub}
\usepackage{mathrsfs}
\usepackage{csquotes}
\newtheorem{theorem}{Theorem}
\newtheorem{definition}{Definition}

\title{Comment on ``Flavor invariants and renormalization- group equations in the leptonic sector with massive Majorana neutrinos''}

\author{Jianlong Lu}

\affiliation{Department of Physics, National University of Singapore,\\
2 Science Drive 3, 117551, Singapore}

\emailAdd{jianlong\_lu@u.nus.edu}

\abstract{Recently in [\emph{JHEP} \textbf{09} (2021) 053], Wang et al. discussed the polynomial ring formed by flavor invariants in the leptonic sector with massive Majorana neutrinos. They have explicitly constructed the finite generating sets of the polynomial rings for both two-generation scenario and three-generation scenario. However, Wang et al.'s claim of the finiteness of the generating sets of the polynomial rings and their calculation by the approach of Hilbert series with generalized Molien-Weyl formula are both based on their assertion that the unitary group ${\rm U}(n,\mathbb{C})$ is reductive, which is unfortunately incorrect. The property of being reductive is only applicable to linear algebraic groups. And it is well-known that the unitary group ${\rm U}(n,\mathbb{C})$ is not even a linear algebraic group. In this paper, we point out the above issue and provide a solution to fill in the accompanying logical gaps in [\emph{JHEP} \textbf{09} (2021) 053]. Some important results from the theory of linear algebraic group, the invariant theory of square matrices and group theory are needed in the analysis. We also clarify some somewhat misleading or vague statements in [\emph{JHEP} \textbf{09} (2021) 053] about the scope of flavor invariants. Note that, although built from incorrect assertion, Wang et al.'s calculation results in [\emph{JHEP} \textbf{09} (2021) 053] are nonetheless correct, which is ultimately because the ring of invariants of ${\rm U}(n,\mathbb{C})$ is isomorphic to that of ${\rm GL}(n,\mathbb{C})$ which is itself reductive. } 
\keywords{Flavor Invariants, Linear Algebraic Groups, Invariant Theory, Majorana Neutrinos}

\arxivnumber{2110.08210}

\begin{document}
\maketitle
\flushbottom

\section{Introduction}
The physical framework of our discussion is encoded in the following effective Lagrangian density for leptons,
\begin{align}
      \mathcal{L}_{{\rm lepton}} = -\overline{l_{{\rm L}}} M_{l} l_{{\rm R}} - \frac{1}{2} \overline{\nu_{{\rm L}}} M_{\nu} \nu_{{\rm L}}^{{\rm C}} + \frac{g}{\sqrt{2}} \overline{l_{{\rm L}}} \gamma^{\mu} \nu_{{\rm L}} W_{\mu}^{-} + {{\rm h.c.}}\ ,
\end{align}
where $M_{l}$ and $M_{\nu}$ are the charged lepton mass matrix and the Majorana neutrino mass matrix respectively. These two mass matrices can be diagonalized by three unitary matrices $\{V_{l}, V_{l}' ,V_{\nu}\}$ via
\begin{align}
      \hat{M}_{l} &= V_{l}^{\dagger} M_{l}  V_{l}' = {\rm Diag}\{m_{e},m_{\mu},m_{\tau}\}, \\
      \hat{M}_{\nu} &= V_{\nu}^{\dagger} M_{\nu}  V_{\nu}^{*} = {\rm Diag}\{m_{1},m_{2},m_{3}\}.
\end{align}
From $V_{l}$ and $V_{\nu}$, one has the Pontecorvo-Maki-Nakagawa-Sakata matrix \cite{pon,mns}:
\begin{align}
     V = V_{l}^{\dagger} V_{\nu}.
\end{align}
The two mass matrices $M_{l}$ and $M_{\nu}$ are not entirely physical, in the sense that one can perform the following transformations without altering the lepton Lagrangian density $\mathcal{L}_{{\rm lepton}}$,
\begin{align}
     l_{{\rm L}} \rightarrow l_{{\rm L}}' = U_{{\rm L}} l_{{\rm L}},\quad \nu_{{\rm L}} \rightarrow \nu_{{\rm L}} ' = U_{{\rm L}} l_{{\rm L}},\quad  l_{{\rm R}} \rightarrow l_{{\rm R}}' = U_{{\rm R}} l_{{\rm R}},
\end{align}
\begin{align}
    M_{l} \rightarrow M_{l}' = U_{{\rm L}} M_{l} U_{{\rm R}}^{\dagger},\quad M_{\nu} \rightarrow  M_{\nu}' =  U_{{\rm L}}  M_{\nu} U_{{\rm L}}^{{\rm T}},
\end{align}
where $U_{{\rm L}}$ and $U_{{\rm R}}$ are arbitrary $3\times 3$ unitary matrices. One can easily eliminate the troublesome factors $U_{{\rm R}}^{\dagger}$ and $U_{{\rm L}}^{{\rm T}}$ by defining the following Hermitian matrices (with $n\in\mathbb{N}^{+}$)
\begin{align}
     H_{l} \equiv M_{l} M_{l}^{\dagger},\quad H_{\nu} \equiv M_{\nu} M_{\nu}^{\dagger},\quad G_{l\nu}^{(n)} \equiv M_{\nu} (H_{l}^{*})^{n} M_{\nu}^{\dagger},
\end{align}
whose transformation rules are easily obtained:
\begin{align}
     H_{l} \rightarrow H_{l}' = U_{{\rm L}} H_{l} U_{{\rm L}}^{\dagger},\quad H_{\nu} \rightarrow H_{\nu}' = U_{{\rm L}} H_{\nu} U_{{\rm L}}^{\dagger},\quad G_{l\nu}^{(n)} \rightarrow G_{l\nu}^{(n)}\ ' = U_{{\rm L}} G_{l\nu}^{(n)} U_{{\rm L}}^{\dagger}.
\end{align}
The satisfying layout of $U_{{\rm L}}$ and $U_{{\rm L}}^{\dagger}$ in the above transformation rules naturally motivates us to write down a class of flavor invariants, ${\rm Tr}(A_{1}A_{2}...A_{k})$, in which $A_{1}, A_{2}, ..., A_{k}\in \{ H_{l}, H_{\nu}, G_{l\nu}^{(n)}\}$ with some positive integer $k$. Of course, such form is not a compulsory requirement for flavor invariants of $\{ H_{l}, H_{\nu}, G_{l\nu}^{(n)}\}$. For example, it is straightforward to prove that ${\rm Det}(G_{l\nu}^{(1)}+H_{l}H_{\nu})$ is also a flavor invariant. Although $n$ can be any positive integer, the Cayley-Hamilton theorem tells us that we only need to focus on $n<N$ when dealing with $N$-generation scenario.\\
In \cite{jar1,jar2}, Jarlskog has given an invariants named after her, which in three-generation scenario can be written as $\mathcal{J} = \cos\theta_{12} \sin\theta_{12} \cos^{2}\theta_{13} \sin\theta_{13} \cos\theta_{23}\sin\theta_{23}\sin\delta$ with parameters in the so-called standard parametrization of mixing matrix. One can use invariants under flavor basis transformations                to construct sufficient and necessary conditions for CP conservation in the leptonic sector within specific ranges of parameters \cite{branco,yz1,lu,yz2}. In \cite{jm}, Jenkins and Manohar analyzed the flavor invariants from the perspective of invariant theory. In \cite{wyz}, Wang et al. claimed that the rings of flavor invariants in the leptonic sector with two and three generations of charged leptons and massive Majorana neutrinos are both finitely generated. For both two-generation and three-generation scenarios, they calculated the numbers of elements in the generating sets, which were also explicitly constructed in \cite{wyz}. However, Wang et al.'s claim of the finiteness of the generating sets and their calculation by the approach of Hilbert series with generalized Molien-Weyl formula are both based on their assertion that the unitary group ${\rm U}(n,\mathbb{C})$ is reductive. This assertion is obviously incorrect, since ${\rm U}(n,\mathbb{C})$ is not even a linear algebraic group. The property of being reductive is only applicable to linear algebraic groups. Such serious gaps in \cite{wyz} imply that the correctness of their results is unjustified. It is the aim of this paper to fill in those gaps of \cite{wyz}.   \\
The remaining part of this paper is organized as follows. In section \ref{s2}, we briefly talk about the notions of linear algebraic group and reductive group. In section \ref{s3}, we present some relevant results in the invariant theory of square matrices, especially for the unitary group ${\rm U}(n,\mathbb{C})$ and the general linear group  ${\rm GL}(n,\mathbb{C})$. Eventually, in section \ref{s4}, we fill in the logical gaps of \cite{wyz}. In section \ref{s5}, we clarify some somewhat misleading or vague statements in \cite{wyz} about the scope of flavor invariants. In section \ref{s6}, we give some remarks on the relation between ${\rm U}(n,\mathbb{C})$ and ${\rm GL}(n,\mathbb{C})$. The readers who are familiar with basic knowledges of algebraic geometry, linear algebraic groups and invariant theory may skip directly to section \ref{s4} and section \ref{s5}.

\section{Linear algebraic groups and reductive groups}
\label{s2}
To fully understand the notion of linear algebraic group requires some knowledges of algebraic geometry. However, it is practically impossible and unnecessary to make this section a comprehensive survey of algebraic geometry or even if only linear algebraic group. Thus only the results needed for our reasoning are presented. For more systematic discussion of linear algebraic group, interested readers are recommended to refer to \cite{borel,gtm21,springer} and the third part of \cite{mac}.\\
First we would like to give the definition of linear algebraic group (see for example section 1 in chapter I of \cite{borel}, section 7 in chapter II of \cite{gtm21}, chapter 2 of \cite{springer} and appendix A of \cite{derksen}):  
\begin{definition}
An algebraic group is an algebraic variety $G$ equipped with morphisms of varieties $\mu: G\times G\rightarrow G$ ($\mu(x,y)=xy$) and $\iota: G\rightarrow G$ ($\iota(x)=x^{-1}$) that give $G$ the structure of a group. If the underlying variety is affine, then $G$ is called a linear algebraic group.
\end{definition}
To be more accessible, a linear algebraic group is a subgroup of the general linear group ${\rm GL}(n,K)$ over some algebraically closed field $K$, defined by polynomial equations. Here we would like to emphasize the keyword ``polynomial''. Some common examples of linear algebraic group are as follows: 
\begin{itemize}
    \item General linear group ${\rm GL}(n,K)$: the space of all $n\times n$ matrices with entries in $K$ and nonzero determinant. This example is not totally trivial, since the definition of ${\rm GL}(n,K)$ is usually formulated with inequality instead of polynomial equation. This issue can be easily dealt with by matching the element $(x_{ij})$ of ${\rm GL}(n,K)$ with the point $(x_{ij},z)\in K^{n^{2}+1}$ satisfying $z\det(x_{ij}) = 1$ \cite{mac}. Note that here we have employed the existence and uniqueness of multiplicative inverse of any element in the field $K$.
    \item Special linear group ${\rm SL}(n,K)$: the space of all $x$ in ${\rm GL}(n,K)$ satisfying ${\rm Det}(x) =1$.
    \item Orthogonal group ${\rm O}(n,K)$: the space of all $x$ in ${\rm GL}(n,K)$ satisfying $x^{\rm T} x = I_{n}$.
    \item Special orthogonal group ${\rm SO}(n,K)$: the intersection of ${\rm SL}(n,K)$ and ${\rm O}(n,K)$.
    \item Symplectic group ${\rm Sp}(2n,K)$: the space of all $x$ in ${\rm GL}(n,K)$ satisfying $x^{\rm T} j x = j$, in which $j\equiv \begin{pmatrix} 0 & I_{n}\\ -I_{n} & 0\end{pmatrix}$.
    \item The group of upper triangular matrices: the space of all $x$ in ${\rm GL}(n,K)$ satisfying $x_{ij}=0$ for $i>j$.
    \item The group of upper unipotent matrices: the space of all $x$ in ${\rm GL}(n,K)$ satisfying $x_{ij}=0$ for $i>j$ and $x_{kk}=1$ for $1\leq k\leq n$.
\end{itemize}
One can see that all of the above examples are defined by polynomial equations of matrix entries. For comparison, one may think about the unitary group ${\rm U}(n,\mathbb{C})$, which is a typical non-example of linear algebraic group. Recall that ${\rm U}(n,\mathbb{C})$ is defined to include any element in ${\rm GL}(n,\mathbb{C})$ whose inverse is its conjugate transpose. It is straightforward to see that the unitary group is not a linear algebraic group, since the definition of unitary group relies on the \emph{complex conjugate} of matrix entries, which is not something polynomials can handle. This fact is also explicitly mentioned by Procesi in section 11 of \cite{procesi}.\\
Another concept worth mentioning is the unipotent radical of a linear algebraic group $G$, which is defined as follows (see for example section 11 in chapter IV of \cite{borel}, section 19.5 in chapter VII of \cite{gtm21}, chapter 6 of \cite{springer} and section 2.2.2 and appendix A.3 of \cite{derksen}):
\begin{definition}
The unipotent radical $R_{u}(G)$ of $G$ is the unique maximal connected unipotent normal subgroup of $G$.
\end{definition}
The unipotent radical $R_{u}(G)$ of $G$ contains all unipotent elements in the radical of $G$. For example, the radical of ${\rm GL}(n,K)$ contains all $n\times n$ scalar matrices with entries from $K$. Among these scalar matrices, only the identity matrix is unipotent due to its exclusive characteristic polynomial $(x-1)^{n}=0$. Therefore the unipotent radical of ${\rm GL}(n,K)$ is said to be trivial.\\
With the notion of unipotent radical, one is ready to define reductive group (see for example section 11 in chapter IV of \cite{borel}, section 19.5 in chapter VII of \cite{gtm21}, chapter 6 of \cite{springer} and appendix A of \cite{derksen}):
\begin{definition}
A linear algebraic group $G$ is said to be reductive if its unipotent radical $R_{u}(G)$ is trivial, i.e., $R_{u}(G) = \{e\}$.
\end{definition}
As mentioned above, ${\rm GL}(n,K)$ has trivial unipotent radical, thus being reductive. There are some other common linear algebraic groups which are also reductive, such as special linear group ${\rm SL}(n,K)$, orthogonal group ${\rm O}(n,K)$, special orthogonal group ${\rm SO}(n,K)$ and symplectic group ${\rm Sp}(2n,K)$. One simple non-example of reductive group is the subgroup of ${\rm GL}(n,K)$ consisting of all upper triangular matrices, sometimes denoted by $B(n,K)$, with $n\geq 2$. The unipotent radical of $B(n,K)$ is one of its subgroups, whose elements all have diagonal entries equal to $1$.\\
At the end of this section, we would like to remark that there exist at least three distinct but related concepts, which are ``reductive group'' (sometimes called ``group theoretically reductive group''), ``linear reductive group'' and ``geometrically reductive group''. \\
Reductive group and linear reductive group are linked by the following theorem \cite{nagata}:
\begin{theorem}
\label{t1}
If ${\rm char}(K) = 0$, then a linear algebraic group is reductive if and only if it is linearly reductive.
\end{theorem}
On the other hand, two directions of the following theorem have been proved in \cite{nagata} and \cite{haboush} respectively.
\begin{theorem}
\label{t2}
For any characteristic of $K$, a group is geometrically reductive if and only if it is reductive.
\end{theorem}
For our discussion of flavor invariants in which $K=\mathbb{C}$, we only need to focus on reductive group since ${\rm char}(\mathbb{C}) = 0$. Thus the definitions of linear reductive group and geometrically reductive group are omitted in this paper. Interested readers may refer to section 2.2.1 and section 2.2.2 of \cite{derksen}.\\
For the benefit of the readers, we would like to remark that, in the vast literature of mathematics, the phrase ``reductive group'' is used by some mathematicians in different circumstances and contexts with different meanings. For example, in Dieudonné and Carrell’s \cite{carrell}, ${\rm U}(n,\mathbb{C})$, as a compact Lie group with all finite dimensional linear representations being completely reducible, is said to be ``reductive''. However, in this case, the ``reductiveness'' of ${\rm U}(n,\mathbb{C})$ is no longer a sufficient condition for the relevant ring of invariants to be finitely generated over $\mathbb{C}$ for rational group action. As explicitly expressed in the corollary of the theorem in section 1 of chapter 3 in \cite{carrell}, there is one more important condition that should also be included to ensure the finiteness. We restate it here for the convenience of interested readers. Let $R = K[a_{1},...,a_{n}]$ be a finitely generated algebra over an arbitrary field $K$ (here we are interested in the case $K = \mathbb{C}$) and $\Gamma$ be a group of algebra automorphisms of $R$ (here we are interested in the case $\Gamma = {\rm U}(n,\mathbb{C})$). Then the condition needed for the subalgebra $I_{\Gamma}\subset R$ (here we are interested in the ring of ${\rm U}(n,\mathbb{C})$ invariants of $\{ H_{l}, H_{\nu}, G_{l\nu}^{(k)}\}$) to be finitely generated over $K$ is that the orbit under $\Gamma$ of each $f\in R$ is contained in a finite dimensional subspace of $R$ over $K$ \cite{carrell}. This condition is not a trivial one and cannot be directly absorbed in the condition that every finite dimensional rational representation of the group of our interests (for example ${\rm U}(n,\mathbb{C})$) is completely reducible. One can also find the relevant description in section 1 of Nagata’s famous \cite{nagata2}. Needless to say, it is not appropriate to choose $R$ to be the ring of ${\rm U}(n,\mathbb{C})$ invariants of $\{ H_{l}, H_{\nu}, G_{l\nu}^{(k)}\}$, otherwise there will be a circular argument. On the other hand, when we work in the context of algebraic group, Hilbert’s finiteness theorem can be neatly formulated, as explicitly stated in the next section. Therefore, with rational group action, when we say that the (linear) reductiveness of a group can alone imply the finiteness of the generating set of a specific ring of invariants, it is the (linearly) reductive group in the context of algebraic group that we are talking about.

\section{Invariant theory of square matrices}
\label{s3}
Invariant theory has a long history of at least one and a half century, with a vast ocean of literature. Two of the most important papers are Hilbert's \cite{hilbert1} and \cite{hilbert2}, which eventually lead to Hilbert's finiteness theorem. For the development of Hilbert’s finiteness theorem, Popov has made a concise summary in his \cite{popov}, which is directly quoted here for the convenience of interested readers:
\begin{displayquote}
The first fundamental theorem was proved by Gordan in 1868 for invariants of binary forms \cite{gordan} (and in 1870 for systems of binary forms \cite{gordan2}). In 1890 Hilbert gave a general nonconstructive proof for any system of forms in any number of variables \cite{hilbert1}. Although in his version he deals with special representations of GL, the idea of the proof can be carried over directly to the general case. H. Weyl \cite{weyl2} accomplished this for any representation of a complex semisimple group, and Mumford \cite{mum} extended this result to the general case of a reductive algebraic group acting regularly on an affine algebraic variety over a field $k$ of characteristic zero. Finally, through the joint efforts of Nagata \cite{nagata2}, Haboush \cite{haboush}, and Mumford \cite{mum}, the restriction on the characteristic of $k$ was removed (in the case ${\rm char}\ k \neq 0$ Hilbert's approach does not work). In this final form the theorem is now called Hilbert's theorem on invariants.
\end{displayquote}
From the above description one can clearly see the special role of algebraic groups in Hilbert’s finiteness theorem.\\
We have no plan to give a detailed review for invariant theory in this section. Only the results needed for our reasoning are presented. Interested readers are recommended to refer to \cite{procesi2} for more systematic discussion of invariant theory of matrices.\\
First we would like to introduce the celebrated Hilbert's finiteness theorem (see for example section 2.2.1 of \cite{derksen} or section 2 of \cite{derksen2}):
\begin{theorem}
\label{t5}
If $G$ is a linearly reductive group and $V$ is a rational representation, then $K[V]^{G}$ is finitely generated over $K$.
\end{theorem}
Next we would like to draw readers' attention to the following theorem (see section 11 of \cite{procesi} for more discussions):
\begin{theorem}
\label{t3}
The unitary invariants of $j$ complex matrices $\{x_{1},..., x_{j}\}$ are polynomials in the elements ${\rm Tr}(w_{i_{1}}... w_{i_{s}})$ where $w_{i}=x_{i}$ or $w_{i}=x_{i}^{\dagger}$. As a ring, it is isomorphic to the ring of ${\rm GL}(n,\mathbb{C})$ invariants of $2j$ matrices $\{x_{1},..., x_{j}, x_{1}^{\dagger},..., x_{j}^{\dagger}\}$.
\end{theorem}
This theorem serves as a bridge between ${\rm U}(n,\mathbb{C})$ invariants and ${\rm GL}(n,\mathbb{C})$ invariants of square matrices. As mentioned in section \ref{s2}, it makes no sense to talk about whether ${\rm U}(n,\mathbb{C})$ is reductive or not. But ${\rm GL}(n,\mathbb{C})$ is indeed (linear) reductive. Now with theorem \ref{t3}, we can play a home game with our familiar (linear) reductive groups instead of an away game with ${\rm U}(n,\mathbb{C})$.\\
The last gear worth mentioning is the theorem given below (see section 1 of \cite{procesi}), which is applicable to ${\rm GL}(n,\mathbb{C})$ invariants:
\begin{theorem}
\label{t4}
Any ${\rm GL}(n,\mathbb{C})$ invariant of $j$ $n\times n$ matrices $\{x_{1},..., x_{j}\}$ is a polynomial in the invariants ${\rm Tr}(x_{i_{1}}... x_{i_{s}})$, in which $x_{i_{1}}... x_{i_{s}}$ runs over all possible noncommutative monomials.
\end{theorem}

\section{Filling in the gaps}
\label{s4}
In section \ref{s2}, we have talked about the notions of linear algebraic group and reductive group. Since the unitary group ${\rm U}(n,\mathbb{C})$ is not even a linear algebraic group, it makes no sense to say that ${\rm U}(n,\mathbb{C})$ is reductive, as wrongly asserted in \cite{wyz}. Now with all necessary tools in our hands, we are well prepared to fill in the gaps in \cite{wyz}. The following discussion is for $n=3$, i.e., the three-generation scenario. When $n=2$, the corresponding discussion is almost the same, thus omitted for the sake of simplicity.\\
In our discussion of flavor invariants, the relevant algebraically closed field $K$ is our familiar $\mathbb{C}$ with characteristic $0$. As mentioned previously, theorem \ref{t1} and theorem \ref{t2} ensure that three concepts ``reductive group'', ``linear reductive group'' and ``geometrically reductive group'' coincide when $K = \mathbb{C}$.\\
Making use of theorem \ref{t3}, one can go over the obstacle caused by the fact that ${\rm U}(3,\mathbb{C})$ is not a reductive group. The unitary invariants of the Hermitian matrices $\{H_{l},H_{\nu},G_{l\nu}^{(1)}, G_{l\nu}^{(2)}\}$ are polynomials in the elements ${\rm Tr}(w_{i_{1}}... w_{i_{s}})$ where $w_{i}\in \{H_{l},H_{\nu},G_{l\nu}^{(1)}, G_{l\nu}^{(2)}\}$. These invariants together form a ring, which is isomorphic to the ring of ${\rm GL}(3,\mathbb{C})$ invariants of the Hermitian matrices $\{H_{l},H_{\nu},G_{l\nu}^{(1)}, G_{l\nu}^{(2)}\}$. In other words, these two rings are indistinguishable from each other. This is consistent with theorem \ref{t4} with $n=3$.\\
As mentioned in section \ref{s2}, ${\rm GL}(3,\mathbb{C})$ is reductive. Then according to theorem \ref{t1} and due to ${\rm char}(\mathbb{C}) = 0$, we know that ${\rm GL}(3,\mathbb{C})$ is also linear reductive. Therefore, theorem \ref{t5} immediately implies that the ring of ${\rm GL}(3,\mathbb{C})$ invariants of the Hermitian matrices $\{H_{l},H_{\nu},G_{l\nu}^{(1)}, G_{l\nu}^{(2)}\}$ is finitely generated over $\mathbb{C}$. Now we can finally say that the ring of the ${\rm U}(3,\mathbb{C})$ invariants of the Hermitian matrices $\{H_{l},H_{\nu},G_{l\nu}^{(1)}, G_{l\nu}^{(2)}\}$ is finitely generated over $\mathbb{C}$.\\
From the above discussion, we can see that the finiteness of the generating set of the ring of the ${\rm U}(3,\mathbb{C})$ invariants of the Hermitian matrices $\{H_{l},H_{\nu},G_{l\nu}^{(1)}, G_{l\nu}^{(2)}\}$ is not a trivial result. Hilbert's finiteness theorem alone is not enough to imply this conclusion. Wang et al.'s statement in \cite{wyz} that ``\emph{This conclusion is a simple and direct consequence of the general mathematical theorem in the invariant theory for the reductive group U(N)}'' is obviously not correct.             \\
The last point we would like to address in this section is the application of Molien-Weyl formula \cite{molien} to calculate Hilbert series in \cite{wyz}. As repeatedly mentioned by Derksen and Kemper in \cite{derksen}, the approach of calculating Hilbert series for infinite groups using generalized Molien-Weyl formula is for (linear) reductive groups. Thus Wang et al.'s calculation in section 3, section 4 and appendix B.3 of \cite{wyz} based on incorrect assumption ``${\rm U}(n,\mathbb{C})$ is reductive'' is unjustified. Fortunately, this problem is not very severe, due to the coincidental fact that ${\rm U}(n,\mathbb{C})$ is the maximal compact subgroup of ${\rm GL}(n,\mathbb{C})$. In the previous analysis, we have transferred our focus from ${\rm U}(n,\mathbb{C})$ invariants to ${\rm GL}(n,\mathbb{C})$ invariants. Because ${\rm GL}(n,\mathbb{C})$ is (linear) reductive, we can calculate Hilbert series using generalized Molien-Weyl formula (see for example section 4.6 of \cite{derksen})
\begin{align}
\label{mw}
    \mathscr{H} (q) = \int [\mathrm{d}\mu]_{G} \frac{1}{{\rm Det}(I - qR(g)) },
\end{align}   
in which $G= {\rm GL}(n,\mathbb{C})$. Due to the non-compactness of ${\rm GL}(n,\mathbb{C})$, the integral in equation \ref{mw} should be performed on the maximal compact subgroup of ${\rm GL}(n,\mathbb{C})$ \cite{derksen,weyl}, which happens to be ${\rm U}(n,\mathbb{C})$.

\section{Some remarks on the scope of flavor invariants}
\label{s5}
A careful reader may have already noticed a problem: from the dimensionless nature of the Jarlskog invariant $\mathcal{J} = \cos\theta_{12} \sin\theta_{12} \cos^{2}\theta_{13} \sin\theta_{13} \cos\theta_{23}\sin\theta_{23}\sin\delta$, it is obvious that the Jarlskog invariant cannot be written as a polynomial of the $34$ basic flavor invariants constructed in \cite{wyz}. Indeed, the Jarlskog invariant is not the only invariant quantity under the flavor basis transformations that cannot be expressed as a polynomial of the $34$ basic flavor invariants constructed in \cite{wyz}. In the three-generation scenario with Majorana neutrinos, there are totally $12$ physical observables in the leptonic sector, including three masses of charged leptons $\{m_{e}, m_{\mu}, m_{\tau}\}$, three masses of Majorana neutrinos $\{m_{1}, m_{2}, m_{3}\}$, three mixing angles $\{\theta_{12}, \theta_{13}, \theta_{23}\}$, one Dirac CP-violating phase $\{\delta\}$ and two Majorana CP-violating phases $\{\rho, \sigma\}$. All of these physical observables are invariants under the flavor basis transformations. But none of them can be written as a polynomial of the basic flavor invariants constructed in \cite{wyz}, which is obvious from their dimensions.\\
The above observation seems to contradict some relevant statements in \cite{wyz} by Wang et al. themselves, such as ``\emph{Any flavor invariants can be expressed as the polynomials of those 34 basic invariants in the generating set.}'' This is due to the sloppy usage of the terminology ``flavor invariant'' in \cite{wyz}. Actually, in the invariant theory of $n\times n$ matrices, the concept of ``invariant'' is rigorously defined \cite{procesi,procesi2}. The relation between the polynomial invariants of a given set of Hermitian matrices under certain transformation and other quantities invariant under the same transformation is very easy to understand. In the case of algebraically closed fields with characteristic $0$ such as $\mathbb{C}$, it has been pointed out as first fundamental theorem in \cite{procesi} and \cite{procesi2} that the ring consisting of the former under ${\rm GL}(n,\mathbb{C})$ is generated by the elements ${\rm Tr}(M)$ with $M$ being any monomial in the given set of Hermitian matrices. And we know that the ring consisting of the former under ${\rm U}(n,\mathbb{C})$ is also generated by the elements ${\rm Tr}(M)$ with $M$ being any monomial in the given set of Hermitian matrices, due to the isomorphism between the ring of ${\rm U}(n,\mathbb{C})$ invariants of $j$ complex square matrices $\{x_{1},...,x_{j}\}$ and the ring of ${\rm GL}(n,\mathbb{C})$ invariants of $2j$ matrices $\{x_{1},...,x_{j},x_{1}^{\dagger},...,x_{j}^{\dagger}\}$. Then from the nontrivial mass dimensions of elements in $\{H_{l},H_{\nu},G_{l\nu}^{(k)} \}$ and the dimensionless nature of Jarlskog invariant, one can immediately see that Jarlskog invariant is not a flavor invariants of $\{H_{l},H_{\nu},G_{l\nu}^{(k)} \}$ under unitary transformation and cannot be expressed as a polynomial of the basic invariants, although Jarlskog invariant is indeed unchanged under such unitary transformation. Therefore there exists no contradiction between the fact that the Jarlskog invariant cannot be written as a polynomial of the $34$ basic flavor invariants constructed in \cite{wyz} and the statement that any flavor invariant can be expressed as a polynomial of those $34$ basic invariants. That is why we say that it is somewhat misleading or vague for Wang et al. to call the Jarlskog invariant a ``flavor invariant'' in the context of invariant theory of square matrices. Similar explanations can be applied to those $12$ physical observables about why they are not invariants of $\{H_{l},H_{\nu},G_{l\nu}^{(1)}, G_{l\nu}^{(2)}\}$ under the flavor basis transformations although being invariant under these transformations.

\section{Some remarks on the relation between ${\rm U}(n,\mathbb{C})$ and ${\rm GL}(n,\mathbb{C})$}
\label{s6}
${\rm U}(n,\mathbb{C})$ is a typical example of Lie groups, and ${\rm GL}(n,\mathbb{C})$ is a typical example of algebraic groups. Lie groups and algebraic groups are characterized by different extra structures equipped on the underlying abstract groups. In \cite{lu2}, we have detailedly proved that ${\rm U}(n,\mathbb{C})$ and ${\rm GL}(n,\mathbb{C})$ are not isomorphic to each other in both the context of Lie group for arbitrary positive integer $n$ and the context of abstract group for arbitrary integer $n\geq 3$. There indeed exist intrinsic differences between ${\rm U}(n,\mathbb{C})$ and ${\rm GL}(n,\mathbb{C})$. These subtleties set up warning signs when one is trying to blindly apply conclusions from one to the other.\\
Nonetheless, one can build a bridge between Lie groups and algebraic groups, such as complexification. A well-known result is that via complexification of compact Lie groups there is a one-to-one correspondence between compact Lie groups (up to differentiable isomorphism) and reductive complex algebraic groups (up to polynomial isomorphism) (see for example chapter 5 of \cite{vinberg}). One example of such correspondence is ${\rm U}(n,\mathbb{C})\rightarrow{\rm GL}(n,\mathbb{C})$. It is the complexification of a compact Lie group that enjoys the nice properties of algebraic groups, instead of the complex Lie group itself. That is why Bröcker and tom Dieck mentioned in their \cite{dieck} about looking at the complexification $G_{\mathbb{C}}$ of a compact Lie group $G$ as an algebraic group.

\section{Summary}
We point out the logical gaps in  [\emph{JHEP} \textbf{09} (2021) 053], which comes from Wang et al.'s incorrect assertion that ${\rm U}(n,\mathbb{C})$ is reductive. We detailedly fill in the gaps with some important results from the theory of linear algebraic group, the invariant theory of square matrices and group theory. Some remarks on the scope of flavor invariants and the relation between ${\rm U}(n,\mathbb{C})$ and ${\rm GL}(n,\mathbb{C})$ are also given.

\section*{Acknowledgements}
The author would like to thank the anonymous referee for his/her helpful suggestions.

\end{document}